\documentclass[%
superscriptaddress,
nofootinbib,
twocolumn,
amsmath,amssymb,
aps,
prd,
a4paper
]{revtex4-2}

\usepackage[colorlinks=true,linkcolor=blue,citecolor=blue,linktocpage=false]{hyperref}

\usepackage{newtxtext,newtxmath}
\usepackage{graphicx}
\usepackage{dcolumn}
\usepackage{bm}
\usepackage{color}

\newcommand{\VEC}{\bm}

\newcommand{\hMpc}{\,h^{-1}\,{\rm Mpc}}

\newcommand{\hk}{\,h\,{\rm Mpc^{-1}}}

\newcommand{\PP}{\bm{\Psi}}

\begin{document}

\preprint{}

\title{Discreteness Effects in the Post-Reconstruction Galaxy Power Spectrum}

\author{Naonori Sugiyama}
\email{nao.s.sugiyama@gmail.com}
\affiliation{National Astronomical Observatory of Japan, Mitaka, Tokyo 181-8588, Japan}

\date{\today}

\begin{abstract}

Recent studies have increasingly recognized the value of analyzing the post-reconstruction galaxy power spectrum for investigations into redshift space distortion (RSD) effects. In this paper, we present a novel theoretical model for the post-reconstruction galaxy power spectrum, designed for RSD analyses. In particular, we emphasize the importance of accounting for discrete effects arising from the reconstruction displacement vector, which have been overlooked. We specifically calculate these discrete effects, i.e., shot noise terms, within the framework of standard perturbation theory at the 1-loop level. In addition, we adopt a formulation that accounts for infrared (IR) effects to accurately model the non-linear damping of the Baryon Acoustic Oscillation (BAO) signal. Our model comprehensively integrates key physical phenomena relevant to the post-reconstruction galaxy power spectrum, such as gravitational non-linearities, RSD effects, bias effects, reconstruction effects, reconstruction-specific shot-noise effects, and the non-linear damping of the reconstructed BAO signal, thereby making it applicable to post-reconstruction RSD analyses.

\end{abstract}

\maketitle

\section{Introduction}

The discovery of new physics beyond the Standard Cosmological Model is now the ultimate goal of cosmology. To this end, reducing errors in measured cosmological statistics is one of the most crucial endeavors in cosmological data analysis. In the analysis of data using spectroscopic galaxies, statistical errors decrease as the survey volume and galaxy number density increase. For this reason, galaxy survey projects are becoming increasingly large-scale, such as the Dark Energy Spectroscopic Instrument~\citep[DESI;][]{DESI:2016fyo}\footnote{\url{http://desi.lbl.gov/}}, the Subaru Prime Focus Spectrograph ~\citep[PFS;][]{PFSTeam:2012fqu}\footnote{\url{https://pfs.ipmu.jp/index.html}}, and Euclid~\citep{EUCLID:2011zbd}\footnote{\url{www.euclid-ec.org}}.

In theoretical aspects, it is well known that using information down to smaller scales reduces statistical errors. Therefore, various improved perturbation theories have been proposed and applied to the analysis of actual observed galaxy data~\cite{Crocce:2005xy,Taruya:2010mx,Baumann:2010tm,Carrasco:2012cv,Carlson:2012bu,Wang:2013hwa}, beyond standard perturbation theory~\cite[SPT;][]{Bernardeau:2001qr}. In recent years, theoretical calculations using emulators, which have been trained with results from $N$-body simulations, are also being conducted~\cite{Nishimichi:2018etk,Kobayashi:2020zsw,Kobayashi:2021oud}.

At small scales, however, since non-Gaussian errors become dominant within the power spectrum covariance matrix, the cumulative signal-to-noise ratio (SNR) of the power spectrum tends to be flat (e.g.,~\cite{Takahashi:2009bq}), meaning that there is no dramatic increase in cosmological information even at small scales. This is thought to be due to the cosmological information of the power spectrum leaking into higher-order statistics, such as the bispectrum, as a result of non-linear gravitational growth. Therefore, the joint analysis of power spectra and bispectra has been actively pursued in recent years to obtain more complete information~\cite{Gil-Marin:2016wya,Slepian:2016kfz,Pearson:2017wtw,DAmico:2019fhj,Philcox:2021kcw,Cabass:2022wjy,DAmico:2022gki,Cabass:2022ymb,Sugiyama:2020uil,DAmico:2022osl,Ivanov:2023qzb,Sugiyama:2023tes,Sugiyama:2023zvd}.

An alternative method for extracting information from higher-order statistics is the reconstruction of galaxy distributions~\cite{Eisenstein:2006nk}. This reconstruction technique was originally proposed to enhance the Baryon Acoustic Oscillation~\cite[BAO;][]{Sunyaev:1970eu,Peebles:1970ag} signal. In principle, the post-reconstruction power spectrum can be represented in a form that includes higher-order statistics, such as the bispectrum, thus partially including the effects of higher-order statistics.

An important feature of the galaxy distribution reconstruction is its ability to dramatically reduce non-Gaussian errors and improve statistical accuracy in cosmological statistics by partially removing non-linear effects. Furthermore, when calculating the cumulative SNR of the power spectrum by taking into account the effects of non-Gaussian terms that give rise to off-diagonal elements in the covariance matrix, the post-reconstruction SNR can increase beyond what is expected from Gaussian errors alone, depending on the scale of interest and the degree of reconstruction~\cite{Hikage:2020fte}. In the bispectrum, when we focus on the primordial non-Gaussian signal, which remains unchanged before and after reconstruction, the SNR in the primordial non-Gaussianity after reconstruction can be larger than when only the Gaussian error is considered~\cite{Shirasaki:2020vkk}, as in the case of the power spectrum. These previous works may imply that the reconstruction of galaxy distributions not only reduces non-Gaussian errors but also propagates information from smaller to larger scales.

\citet{Wang:2022nlx} also showed that a joint analysis of the pre- and post-reconstruction power spectra can extract more cosmological information than examining each separately. This enhancement in information extraction is likely attributable to the reduced cross-covariance between the pre- and post-reconstruction power spectra at small scales, leading to a tendency for the information within them to be independent. This independence was empirically validated through simulation-based measurements. Additionally, \citet{Sugiyama:2024eye} found that the correlation between pre- and post-reconstruction fields invariably shows exponential decay due to infrared effects, thereby providing a mathematical foundation for their independence.

Such unique features of the reconstruction have been suggested to contribute not only to the data analysis related to BAO but also to the improvement of various cosmological information, including Redshift Space Distortions~\cite[RSD;][]{Hikage:2020fte,Wang:2022nlx} and primordial non-Gaussianities~\cite{Shirasaki:2020vkk}. However, cosmological analyses using the post-reconstruction power spectrum beyond the BAO analysis have never been performed on actual galaxy data.

The purpose of this paper is to present a theoretical model corresponding to the post-reconstruction power spectrum measured from observed galaxy data, in preparation for future post-reconstruction RSD analyses. To date, theoretical calculations of the post-reconstruction power spectrum that go beyond linear theory have been carried out using one-loop corrections in SPT~\cite{Schmittfull:2015mja,Hikage:2017tmm,Hikage:2019ihj,Sugiyama:2024eye} and using the Zel'dovich approximation~\cite{Padmanabhan:2008dd,Seo:2015eyw,White:2015eaa,Chen:2019lpf}. For example, \citet{Chen:2019lpf} and \citet{Sugiyama:2024eye} include both the RSD effect~\cite{Kaiser:1987qv} and the bias effect~\cite{Desjacques:2016bnm}, for the purpose of explaining the observed spectroscopic galaxy distribution. 

However, to adequately describe the measured post-reconstruction power spectrum, one more crucial effect remains to be accounted for: the discrete effects unique to reconstruction, namely, the shot noise effects. While the discrete effect in the post-reconstruction power spectrum was first pointed out by~\citet{White:2010qd}, this study was limited to the effects appearing in the Gaussian function that damps the BAO signal. In principle, however, the discretization effects due to reconstruction should manifest themselves across the entire range of non-linear effects. In this paper, we formulate the specific form of the shot noise effects in the post-reconstruction one-loop corrections. Consequently, we assemble the remaining piece necessary for comparison with the observed post-reconstruction galaxy power spectrum.

Considering only the 1-loop correction term in SPT is still insufficient to adequately explain the post-reconstruction power spectrum as actually observed. This insufficiency arises from the well-known limitation that the SPT 1-loop correction term does not fully account for the non-linear damping effects of the BAO signal~(e.g., see Figure $1$ in~\cite{Blas:2016sfa}). To address this issue, it is necessary to implement a resummation of infrared (IR) effects~\citep{Zeldovich:1969sb,Eisenstein:2006nj,Crocce:2007dt,Matsubara:2007wj,Sugiyama:2013gza,Senatore:2014via,Baldauf:2015xfa,Blas:2016sfa,Senatore:2017pbn,Ivanov:2018gjr,Lewandowski:2018ywf,Sugiyama:2020uil}; \citet{Sugiyama:2024eye}, in particular, proposes a IR-resummed model of the post-reconstruction galaxy power spectrum, including the 1-loop correction term in SPT.

In this paper, we extend \citet{Sugiyama:2024eye}'s model to include reconstruction-specific discrete effects. This approach allows us to construct a model that is applicable to smaller scales while accounting for the non-linear damping of the BAO. Furthermore, our calculations also show that the shot noise term appearing in the non-linear damping of the post-reconstruction BAO signal reproduces the results obtained by \citet{White:2010qd}.

The paper is organized as follows. In Section~\ref{Sec:DensityFluctuation}, we review the discrete effects arising from the density fluctuations before reconstruction. In particular, we give the discrete representation in higher-order statistics, i.e., the representation with shot noise terms, which is required for the calculation of the post-reconstruction power spectrum. In Section~\ref{Sec:PostRecon}, we give a discrete representation of the post-reconstruction power spectrum, and in Section~\ref{Sec:PT}, we specifically calculate shot noise terms in the 1-loop corrections in the framework of perturbation theory. Section~\ref{Sec:IRmodel} presents a model of the post-reconstruction power spectrum with resummation of the IR effects. Discussion and conclusions are given in Section~\ref{Sec:Conclusions}.

\section{Pre-reconstruction case}
\label{Sec:DensityFluctuation}

In preparation for the post-reconstruction case, we summarize the discrete effects in terms up to the fourth-order of the density fluctuations, including the power spectrum, bispectrum, and trispectrum.

\subsection{Galaxy density fluctuations}

Following~\cite{Peebles:1980yev}, we divide the space into the infinitesimal grid cells of volume $\delta V$. Each cell is allowed to contain at most one galaxy, which can be described by the occupation number $n_i$ such that $n_i=1$ if the $i$th cell contains a galaxy, and $n_i=0$ otherwise. Therefore, for any natural number $n$, the occupation number satisfies 
\begin{eqnarray}
    n_i = n_i^2 = \dots = n_i^n  \;.
    \label{Eq:n_i}
\end{eqnarray}

The galaxy number density can be represented as
\begin{eqnarray}
    n_{\rm gal}(\VEC{x}) = \sum_i\, n_i\, \delta_{\rm D}(\VEC{x}-\VEC{x}_i)\;,
\end{eqnarray}
where $\VEC{x}_i$ denotes the position of the $i$th galaxy, and $\delta_{\rm D}$ represents the three-dimensional delta function. The occupation number $n_i$ is defined using the galaxy number density as
\begin{eqnarray}
    n_i = n_{\rm gal}(\VEC{x}_i)\delta V\;,
\end{eqnarray}
because the delta function is discretized as
\begin{eqnarray}
    \delta_{\rm D}(\VEC{x}_i-\VEC{x}_j)
    \xrightarrow[\rm discretized]{} \delta_{\rm K}(i,j)\, /\, \delta V \;,
\end{eqnarray}
where $\delta_{\rm K}(i,j)$ represents the Kronecker delta, which equals $1$ when $i=j$ and $0$ otherwise. The total number of galaxies is then calculated by summing over the occupation numbers,
\begin{eqnarray}
    N_{\rm gal} = \int d^3x\, n_{\rm gal}(\VEC{x})
    = \sum_i n_i \;.
\end{eqnarray}
The galaxy mean number density is given by
\begin{eqnarray}
    \bar{n} = \frac{N_{\rm gal}}{V}\;,
\end{eqnarray}
where $V$ is the survey volume.

In a manner analogous to the case with galaxies, the random number density is described by
\begin{eqnarray}
    n_{\rm ran}(\VEC{x}) = \sum_i\, m_i\, \delta_{\rm D}(\VEC{x}-\VEC{x}_i)\;,
\end{eqnarray}
where the $m_i = n_{\rm ran}(\VEC{x}_i) \delta V$ represents the occupation number corresponding to random particles. The total number of random particles is then given by
\begin{eqnarray}
    N_{\rm ran} = \int d^3x\, n_{\rm ran}(\VEC{x})
    = \sum_i m_i \;.
\end{eqnarray}
We define the ratio between $N_{\rm gal}$ and $N_{\rm ran}$ as
\begin{eqnarray}
    \alpha = \frac{N_{\rm gal}}{N_{\rm ran}}\;.
    \label{Eq:alpha}
\end{eqnarray}
To ensure that the discretization effects arising from random particles are sufficiently small compared to those originating from galaxies, the value of $\alpha$ is conventionally chosen to be of the order of $10^{-2}$.
Furthermore, we assume that galaxies and random particles are never in the same position, leading to 
\begin{eqnarray}
    n_i m_i = 0  \;.
    \label{Eq:nm}
\end{eqnarray}

Using the galaxy and random number densities, the galaxy density fluctuation is represented as
\begin{eqnarray}
    \delta(\VEC{x}) 
    &=& \left( 1/\bar{n} \right)\left[ n_{\rm gal}(\VEC{x}) - \alpha n_{\rm ran}(\VEC{x})  \right] \nonumber \\
    &=& \left( 1/\bar{n} \right)\sum_i\, (n_i - \alpha m_i)\, \delta_{\rm D}(\VEC{x}-\VEC{x}_i)\;,
    \label{Eq:delta_discrete}
\end{eqnarray}
In Fourier space, this becomes
\begin{eqnarray}
    \widetilde{\delta}(\VEC{k}) &=&\int d^3x e^{-i\VEC{k}\cdot\VEC{x}} \delta(\VEC{x}) \nonumber \\
    &=&  \left( 1/\bar{n} \right)\sum_i\, (n_i - \alpha m_i)\,  e^{-i\VEC{k}\cdot\VEC{x}_i}\;,
    \label{Eq:delta_Fourier}
\end{eqnarray}
where the tilde denotes a Fourier-transformed quantity.

Throughout the remainder of this paper, for the convenience of subsequent calculations, we introduce
\begin{eqnarray}
    \widetilde{\delta}_i(\VEC{k}) = \left( 1/\bar{n} \right)\, (n_i - \alpha m_i)\,  
    e^{-i\VEC{k}\cdot\VEC{x}_i}\;,
    \label{Eq:delta_i}
\end{eqnarray}
and rewrite the Fourier-transformed density fluctuations as 
\begin{eqnarray}
    \widetilde{\delta}(\VEC{k}) = \sum_i \widetilde{\delta}_i(\VEC{k})\;.
\end{eqnarray}
This notation allows for the explicit description of the discrete effects in density fluctuations. 

\subsection{Power spectrum}

Using the notation presented in Eq.~(\ref{Eq:delta_i}), the power spectrum in the discrete representation is represented as
\begin{eqnarray}
    \left\langle \sum_{i\neq j} \widetilde{\delta}_i(\VEC{k})\widetilde{\delta}_j(\VEC{k}') \right\rangle
    = (2\pi)^3\delta_{\rm D}(\VEC{k}+\VEC{k}') P(\VEC{k}) \;,
    \label{Eq:P}
\end{eqnarray}
where $\langle \cdots \rangle$ denotes the ensemble average. In the left-hand side of the above equation, note that the summation is performed under the condition $i \neq j$. This is because the power spectrum is a measure for computing the correlation between two different galaxies, so it is necessary to avoid counting the same galaxy. 

The effect of counting the same galaxy where $i=j$ is called the shot-noise effect. In the case of $i=j$, we obtain
\begin{eqnarray}
     \sum_i \widetilde{\delta}_i(\VEC{k})\widetilde{\delta}_i(-\VEC{k})
     &=&  \left( 1/\bar{n} \right)^2 \left( 1 + \alpha  \right)\, N_{\rm gal}\;,
     \label{Eq:delta_i_1}
\end{eqnarray}
where we used $n_i m_i = 0$. Since $\alpha$ in Eq.~(\ref{Eq:alpha}) is usually chosen to be sufficiently smaller than $1$, e.g., on the order of $10^{-2}$, neglecting it, this equation can be approximated as
\begin{eqnarray}
    \sum_i \widetilde{\delta}_i(\VEC{k})\widetilde{\delta}_i(-\VEC{k})
    \approx  \left( 1/\bar{n} \right)^2\, N_{\rm gal} \;.
\end{eqnarray}
Consequently, we derive the power spectrum with the shot-noise term as
\begin{eqnarray}
    \left\langle \widetilde{\delta}(\VEC{k})\widetilde{\delta}(\VEC{k}') \right\rangle
    &=& \frac{(2\pi)^3\delta_{\rm D}(\VEC{k}+\VEC{k}')}{V}
    \left\langle \left(\sum_{i\neq j}  + \sum_{i=j} \right)\widetilde{\delta}_i(\VEC{k})\widetilde{\delta}_j(-\VEC{k}) \right\rangle \nonumber \\
    &=& (2\pi)^3\delta_{\rm D}(\VEC{k}+\VEC{k}') P_{\rm N}(\VEC{k})\;,
    \label{Eq:P_NN}
\end{eqnarray}
where
\begin{eqnarray}
    P_{\rm N}(\VEC{k}) = P(\VEC{k}) + \frac{1}{\bar{n}} \;.
    \label{Eq:P_N}
\end{eqnarray}
From Eq.~(\ref{Eq:P_NN}), when density fluctuations are calculated from discrete particles, it is understood that the power spectrum computed from the product of two density fluctuations invariably contains the shot noise term.

\subsection{Bispectrum}
\label{Sec:Bispectra}

Although the main focus of this paper is on the power spectrum, the bispectrum is used to express the post-reconstruction power spectrum in Section~\ref{Sec:PostRecon_P}.

The bispectrum in the discrete representation is represented as
\begin{eqnarray}
    \left\langle\sum_{i\neq j \neq k}
    \widetilde{\delta}_i(\VEC{k}_1)\widetilde{\delta}_j(\VEC{k}_2)\widetilde{\delta}_k(\VEC{k}_3) \right\rangle
    =(2\pi)^3\delta_{\rm D}(\VEC{k}_{123}) 
    B(\VEC{k}_1,\VEC{k}_2,\VEC{k}_3) \;, \nonumber \\
    \label{Eq:B}
\end{eqnarray}
where $\VEC{k}_{123}=\VEC{k}_1+\VEC{k}_2+\VEC{k}_3$.

As in the case of the power spectrum, when calculating the bispectrum from the product of three density fluctuations, shot-noise terms for the bispectrum appear.
\begin{eqnarray}
    \left\langle \widetilde{\delta}(\VEC{k}_1)\widetilde{\delta}(\VEC{k}_2)
    \widetilde{\delta}(\VEC{k}_3) \right\rangle 
    &=& 
    \left\langle \sum_{i,j,k}\widetilde{\delta}_i(\VEC{k}_1)\widetilde{\delta}_j(\VEC{k}_2)
    \widetilde{\delta}_k(\VEC{k}_3) \right\rangle  \nonumber \\
    &=& (2\pi)^3\delta_{\rm D}(\VEC{k}_{123}) 
    B_{\rm N_{\rm all}}(\VEC{k}_1,\VEC{k}_2,\VEC{k}_3) \;.
\end{eqnarray}
To calculate $B_{\rm N_{\rm all}}$, we decompose $\sum_{i,j,k}$ into the following five components:
\begin{eqnarray}
    \sum_{i,j,k} = \sum_{i\neq j \neq k}  + \sum_{i \neq j, i=k} + \sum_{i \neq j, j=k} + \sum_{i \neq k, i=j}  + \sum_{i = j = k} \;.
\end{eqnarray}
The first term on the right-hand side of the above equation leads to the bispectrum, according to the definition in Eq.~(\ref{Eq:B}). To calculate the second term, using Eq.~(\ref{Eq:delta_i}), we derive the following relation for two different wave vectors $\VEC{k}$ and $\VEC{k}'$:
\begin{eqnarray}
    \sum_i \widetilde{\delta}_i(\VEC{k})\widetilde{\delta}_i(\VEC{k}')
    &=&  \left( 1/\bar{n} \right) \widetilde{\delta}(\VEC{k}+\VEC{k}')\nonumber \\
    &+& 
    \left( 1/\bar{n} \right)^2  \alpha \left( 1 + \alpha \right) \widetilde{n}_{{\rm ran}}(\VEC{k}+\VEC{k}') \;.
\end{eqnarray}
Ignoring the term proportional to $\alpha$, this relation can be approximated as
\begin{eqnarray}
    \sum_i \widetilde{\delta}_i(\VEC{k})\widetilde{\delta}_i(\VEC{k}')
    &\approx&  \left( 1/\bar{n} \right) \widetilde{\delta}(\VEC{k}+\VEC{k}')\;.
    \label{Eq:DiDi}
\end{eqnarray}
Using Eq.~(\ref{Eq:DiDi}), the second term is calculated as
\begin{eqnarray}
     \left\langle \sum_{i\neq j,i=k}\widetilde{\delta}_i(\VEC{k}_1)\widetilde{\delta}_j(\VEC{k}_2)
    \widetilde{\delta}_k(\VEC{k}_3) \right\rangle  
    = (2\pi)^3\delta_{\rm D}\left( \VEC{k}_{123} \right)\frac{1}{\bar{n}} P(\VEC{k}_2)\;.
\end{eqnarray}
The third and fourth terms can be similarly calculated. Finally, the fifth term can be calculated as 
\begin{eqnarray}
    \left\langle \sum_i \widetilde{\delta}_i(\VEC{k}_1)\widetilde{\delta}_i(\VEC{k}_2)\widetilde{\delta}_i(\VEC{k}_3) \right\rangle
    &=& (2\pi)^3\delta_{\rm D}(\VEC{k}_{123}) \left(1/\bar{n} \right)^2 \left( 1 - \alpha^2  \right) \nonumber \\
    &\approx& (2\pi)^3\delta_{\rm D}(\VEC{k}_{123})\left(1/\bar{n} \right)^3   \;,
\end{eqnarray}
where $\alpha^2$ is ignored in the final line. Consequantly, we obtain
\begin{eqnarray}
    B_{\rm N_{\rm all}}(\VEC{k}_1,\VEC{k}_2,\VEC{k}_3)
    &=&  B(\VEC{k}_1,\VEC{k}_2,\VEC{k}_3) \nonumber \\
    &+& \frac{1}{\bar{n}}\left[ P(\VEC{k}_1) + P(\VEC{k}_2) + P(\VEC{k}_3) \right]
    + \frac{1}{\bar{n}^2} \;.
    \label{Eq:B_Nall}
\end{eqnarray}

In the computation of the post-reconstruction power spectrum in Section~\ref{Sec:PostRecon_P}, the bispectrum with shot noise terms appears in a form different from $B_{\rm N_{\rm all}}$ in Eq.~(\ref{Eq:B_Nall}). In this paper, we define the bispectrum with shot noise for the following special case as
\begin{eqnarray}
    \left\langle\sum_{i\neq j, k}
    \widetilde{\delta}_i(\VEC{k}_1)\widetilde{\delta}_j(\VEC{k}_2)\widetilde{\delta}_k(\VEC{k}_3) \right\rangle
    =(2\pi)^3\delta_{\rm D}(\VEC{k}_{123}) 
    B_{\rm N_{12}}(\VEC{k}_1,\VEC{k}_2,\VEC{k}_3) \;, \nonumber \\
    \label{Eq:B_NN}
\end{eqnarray}
where
\begin{eqnarray}
    B_{\rm N_{12}}(\VEC{k}_1,\VEC{k}_2,\VEC{k}_3)
    = B(\VEC{k}_1,\VEC{k}_2,\VEC{k}_3)
    + \frac{1}{\bar{n}} \left[ P(\VEC{k}_1) + P(\VEC{k}_2) \right] \;.
    \label{Eq:B_N}
\end{eqnarray}
In this $B_{\rm N_{12}}$, it is important to note that the exchange symmetry among $\VEC{k}_1$, $\VEC{k}_2$, and $\VEC{k}_3$ is no longer preserved. Only $\VEC{k}_1$ and $\VEC{k}_2$ are exchangeable.

\subsection{Product of four density fluctuations, and trispectrum}

In this subsection, we calculate the discrete effects that arise from the product of four density fluctuations. In doing so, Wick's theorem allows for the decomposition into terms consisting of products of power spectra and terms originating from the trispectrum. 

In our notation, the trispectrum is represented as follows:
\begin{eqnarray}
    && \left\langle\sum_{i\neq j \neq k \neq l}
    \widetilde{\delta}_i(\VEC{k}_1)\widetilde{\delta}_j(\VEC{k}_2)
    \widetilde{\delta}_k(\VEC{k}_3)\widetilde{\delta}_l(\VEC{k}_4) \right\rangle_{\rm c} 
    \nonumber \\
    &=& 
    (2\pi)^3\delta_{\rm D}(\VEC{k}_{1234}) 
    T(\VEC{k}_1,\VEC{k}_2,\VEC{k}_3,\VEC{k}_4) \;, \nonumber \\
\end{eqnarray}
where $\langle \cdots \rangle_{\rm c}$ signifies that only the connected part is extracted when calculating the ensemble average.

Similar to the case of the bispectrum in Eq.~(\ref{Eq:B_NN}), this paper considers the following special case.
\begin{eqnarray}
    && \left\langle\sum_{i\neq j, k, l}
    \widetilde{\delta}_i(\VEC{k}_1)\widetilde{\delta}_j(\VEC{k}_2)
    \widetilde{\delta}_k(\VEC{k}_3)\widetilde{\delta}_l(\VEC{k}_4) \right\rangle \nonumber \\
    &=& 
   (2\pi)^3\delta_{\rm D}\left( \VEC{k}_1+\VEC{k}_2 \right) 
   (2\pi)^3\delta_{\rm D}\left( \VEC{k}_3+\VEC{k}_4 \right) P(\VEC{k}_1) P_{\rm N}(\VEC{k}_3) \nonumber \\
    &+& 
   (2\pi)^3\delta_{\rm D}\left( \VEC{k}_1+\VEC{k}_3 \right) 
   (2\pi)^3\delta_{\rm D}\left( \VEC{k}_2+\VEC{k}_4 \right) P_{\rm N}(\VEC{k}_1) P_{\rm N}(\VEC{k}_2) \nonumber \\
    &+& 
   (2\pi)^3\delta_{\rm D}\left( \VEC{k}_1+\VEC{k}_4 \right) 
   (2\pi)^3\delta_{\rm D}\left( \VEC{k}_2+\VEC{k}_3 \right) P_{\rm N}(\VEC{k}_1) P_{\rm N}(\VEC{k}_2) \nonumber \\
    &+& 
    (2\pi)^3\delta_{\rm D}(\VEC{k}_{1234}) 
    T_{\rm N_{12}}(\VEC{k}_1,\VEC{k}_2,\VEC{k}_3,\VEC{k}_4) \;. 
    \label{Eq:T_NN}
\end{eqnarray}
The first term on the right-hand side represents the product of the power spectra with and without the shot noise effect ($P \times P_{\rm N}$). The second and third terms are the products of two $P_{\rm N}$ terms ($P_{\rm N}\times P_{\rm N}$). The last term corresponds to the trispectrum with shot noise terms. By performing calculations similar to those for the bispectrum in Section~\ref{Sec:Bispectra}, we obtain
\begin{eqnarray}
    \hspace{-0.7cm}
    && T_{\rm N_{12}}(\VEC{k}_1,\VEC{k}_2,\VEC{k}_3,\VEC{k}_4) \nonumber \\
    \hspace{-0.7cm}
     &=&
    T(\VEC{k}_1,\VEC{k}_2,\VEC{k}_3,\VEC{k}_4) \nonumber \\
    \hspace{-0.7cm}
 &+& \left( 1/\bar{n} \right) \Big[ 
        B(\VEC{k}_1+\VEC{k}_3,\VEC{k}_2,\VEC{k}_4) + B(\VEC{k}_1+\VEC{k}_4,\VEC{k}_2,\VEC{k}_3)  \nonumber \\
     \hspace{-0.7cm}
     \hspace{-0.7cm}
&& \hspace{0.9cm} + B(\VEC{k}_1,\VEC{k}_2+\VEC{k}_4,\VEC{k}_3) + B(\VEC{k}_1,\VEC{k}_2+\VEC{k}_3,\VEC{k}_4)  \nonumber \\
     \hspace{-0.7cm}
&& \hspace{0.9cm}  + B(\VEC{k}_1,\VEC{k}_2,\VEC{k}_3+\VEC{k}_4) \Big] \nonumber \\
     \hspace{-0.7cm}
&+& \left( 1/\bar{n} \right)^2 \Big[ 
        P(\VEC{k}_1) + P(\VEC{k}_2) 
    + P(\VEC{k}_1+\VEC{k}_3) + P(\VEC{k}_1+\VEC{k}_4) \Big] \;.
    \label{Eq:T_N}
\end{eqnarray}
Here, note that only the exchange of $\VEC{k}_1$ and $\VEC{k}_2$, as well as the exchange of $\VEC{k}_3$ and $\VEC{k}_4$, are symmetric.

\section{Post-reconstruction case}
\label{Sec:PostRecon}

In this section, we calculate discrete effects in the post-reconstruction power spectrum.

\subsection{Galaxy density fluctuations}

To reconstruct the distribution of galaxies, the displacement vector used for reconstruction is derived from the observed galaxy density fluctuations as follows~\cite{Eisenstein:2006nk}:
\begin{eqnarray}
    \VEC{s}(\VEC{x}) = 
    i\,\int \frac{d^3p}{(2\pi)^3} e^{i\VEC{p}\cdot\VEC{x}} 
    \VEC{R}(\VEC{p})\, \widetilde{\delta}(\VEC{p})
    \label{Eq:S}
\end{eqnarray}
with 
\begin{eqnarray}
    \VEC{R}(\VEC{p}) = \left( \frac{\VEC{p}}{p^2} \right)
    \left(  - \frac{W_{\rm G}(pR_{\rm s})}{b_{1, \rm fid}} \right)\;,
\end{eqnarray}
where $b_{1, \rm fid}$ is the fiducial linear bias parameter input for reconstruction, $W_{\rm G}(pR_{\rm s}) = \exp\left( -p^2R_{\rm s}^2/2 \right)$ is a Gaussian filter function, and $R_{\rm s}$ is the input smoothing scale. 

The reconstruction of the galaxy distribution is the operation of moving the positions of galaxies and random particles by using $\VEC{s}(\VEC{x})$. The post-reconstruction density fluctuation is then given by~\cite{Sugiyama:2020uil,Shirasaki:2020vkk}
\begin{eqnarray}
    \delta_{\rm rec}(\VEC{x}) 
    = \int d^3x'\, \delta(\VEC{x}')\, \delta_{\rm D}(\VEC{x} - \VEC{x}' - \VEC{s}(\VEC{x}'))\;.
    \label{Eq:delta_rec}
\end{eqnarray}
Substituting Eq.~(\ref{Eq:delta_discrete}) into Eqs.~(\ref{Eq:S}) and (\ref{Eq:delta_rec}) results in 
\begin{eqnarray}
    \delta_{\rm rec}(\VEC{x})
    = \left( 1/\bar{n} \right)\sum_i\, \left( n_i - \alpha m_i \right)\, 
    \delta_{\rm D}(\VEC{x}-\VEC{x}_i-\VEC{s}_i)\;,
    \label{Eq:delta_rec_discrete}
\end{eqnarray}
where 
\begin{eqnarray}
    \hspace{-0.5cm}
    \VEC{s}_i =  
    i\,\int \frac{d^3p}{(2\pi)^3} e^{i\VEC{p}\cdot\VEC{x}_i} 
    \VEC{R}(\VEC{p})
    \sum_{j}\, \widetilde{\delta}_{j}(\VEC{p})\;.
    \label{Eq:S_i}
\end{eqnarray}
In Fourier space, Eq.~(\ref{Eq:delta_rec_discrete}) becomes
\begin{eqnarray}
    \widetilde{\delta}_{\rm rec}(\VEC{k}) = \sum_i \widetilde{\delta}_i(\VEC{k}) e^{-i\VEC{k}\cdot\VEC{s}_i} \;.
    \label{Eq:delta_rec_F}
\end{eqnarray}

In Section~\ref{Sec:PT}, we will calculate the discretization effect included in the 1-loop correction of the power spectrum. For this purpose, we expand $e^{-i\VEC{k}\cdot\VEC{s}_i}$ in Eq.~(\ref{Eq:delta_rec_F}) to the second order of $\VEC{s}_i$:
\begin{eqnarray}
    \hspace{-0.5cm}
    \widetilde{\delta}_{\rm rec}(\VEC{k}) \approx
    \sum_i \widetilde{\delta}_i(\VEC{k}) 
    \left[ 1 + \left( -i\VEC{k}\cdot\VEC{s}_i \right) 
    + \frac{1}{2}\left( -i\VEC{k}\cdot\VEC{s}_i \right)^2 \right] \;,
\end{eqnarray}
where each term is given by
\begin{eqnarray}
    && \sum_i \widetilde{\delta}_i(\VEC{k}) \left( -i\VEC{k}\cdot\VEC{s}_i \right)\nonumber \\
    &=&\frac{1}{2}\int \frac{d^3p_1}{(2\pi)^3}\int \frac{d^3p_2}{(2\pi)^3} (2\pi)^3\delta_{\rm D}(\VEC{k}-\VEC{p}_1-\VEC{p}_2) \nonumber \\
    &\times&  \sum_{i,j} 
    \Big[ \left[ \VEC{k}\cdot\VEC{R}(\VEC{p}_2) \right]
    \widetilde{\delta}_i(\VEC{p}_1) \widetilde{\delta}_{j}(\VEC{p}_2) 
        \nonumber \\
        && \hspace{0.5cm}
        + \left[ \VEC{k}\cdot\VEC{R}(\VEC{p}_1) \right] 
    \widetilde{\delta}_i(\VEC{p}_2)\widetilde{\delta}_{j}(\VEC{p}_1) \Big]\;,
        \label{Eq:D_S2}
\end{eqnarray}
and
\begin{eqnarray}
    && \frac{1}{2}\sum_i \widetilde{\delta}_i(\VEC{k}) \left( -i\VEC{k}\cdot\VEC{s}_i \right)^2\nonumber \\
    &=&\frac{1}{6} \int \frac{d^3p_1}{(2\pi)^3}\int \frac{d^3p_2}{(2\pi)^3}\int \frac{d^3p_3}{(2\pi)^3}
    (2\pi)^3\delta_{\rm D}(\VEC{k}-\VEC{p}_1-\VEC{p}_2-\VEC{p}_3) \nonumber \\
    &\times&  \sum_{i,j,k} 
    \Big[ \left[ \VEC{k}\cdot\VEC{R}(\VEC{p}_2) \right]\left[ \VEC{k}\cdot\VEC{R}(\VEC{p}_3) \right] \widetilde{\delta}_i(\VEC{p}_1) \widetilde{\delta}_{j}(\VEC{p}_2) \widetilde{\delta}_{k}(\VEC{p}_3) 
        \nonumber \\
        &&\hspace{0.5cm}+
        \left[ \VEC{k}\cdot\VEC{R}(\VEC{p}_1) \right]
        \left[ \VEC{k}\cdot\VEC{R}(\VEC{p}_2)\right] \widetilde{\delta}_i(\VEC{p}_3) \widetilde{\delta}_{j}(\VEC{p}_1) \widetilde{\delta}_{k}(\VEC{p}_2) 
        \nonumber \\
        &&\hspace{0.5cm}+
        \left[ \VEC{k}\cdot\VEC{R}(\VEC{p}_1) \right]\left[ \VEC{k}\cdot\VEC{R}(\VEC{p}_3)\right]
        \widetilde{\delta}_i(\VEC{p}_2) 
        \widetilde{\delta}_{j}(\VEC{p}_1) \widetilde{\delta}_{k}(\VEC{p}_3) 
    \Big] \;.
        \label{Eq:D_S3}
\end{eqnarray}
We can see from Eqs.~(\ref{Eq:D_S2}) and (\ref{Eq:D_S3}) that discrete effects invariably emerge from $\VEC{s}_i$ when calculating post-reconstruction density fluctuations.

\subsection{Power spectrum}
\label{Sec:PostRecon_P}

Using Eq.~(\ref{Eq:delta_rec_F}), the post-reconstruction power spectrum in the discrete representation is expressed as
\begin{eqnarray}
   && \left\langle \sum_{i\neq j}\widetilde{\delta}_i(\VEC{k})\widetilde{\delta}_j(\VEC{k}')
    e^{-i\VEC{k}\cdot\VEC{s}_i}e^{-i\VEC{k}'\cdot\VEC{s}_j} \right\rangle \nonumber \\
   &=&  (2\pi)^3\delta_{\rm D}(\VEC{k}+\VEC{k}') P_{\rm rec}(\VEC{k})\;.
    \label{Eq:P_rec}
\end{eqnarray}
We expand $e^{-i\VEC{k}\cdot\VEC{s}_i}e^{-i\VEC{k}'\cdot\VEC{s}_j}$ in Eq.~(\ref{Eq:P_rec}) up to the second order of $\left[ (-i\VEC{k}\cdot\VEC{s}_i) + (-i\VEC{k}'\cdot\VEC{s}_j) \right]$:
\begin{eqnarray}
    P_{\rm rec}(\VEC{k}) \approx P(\VEC{k}) + P_{s}(\VEC{k}) + P_{s^2}(\VEC{k}) \;,
\end{eqnarray}
where
\begin{eqnarray}
    && \left\langle \frac{1}{n!}\sum_{i \neq j}\widetilde{\delta}_i(\VEC{k}) \widetilde{\delta}_j(\VEC{k}') 
    \left[ \left( -i\VEC{k}\cdot\VEC{s}_i \right) + \left( -i\VEC{k}'\cdot\VEC{s}_j \right) \right]^n \right\rangle \nonumber \\
    &=& (2\pi)^3\delta_{\rm D}(\VEC{k}+\VEC{k}') P_{s^n}(\VEC{k})
\end{eqnarray}
for $n\geq1$.

\subsubsection{$P_{\rm s}$}

Using Eq.~(\ref{Eq:B_N}), $P_{s}(\VEC{k})$ can be calculated as
\begin{eqnarray}
    P_{\rm s}(\VEC{k}) &=&   \int \frac{d^3p}{(2\pi)^3}
    \Big[ \left[ \VEC{k}\cdot\VEC{R}(\VEC{k}-\VEC{p}) \right]B_{\rm N_{12}}(-\VEC{k},\VEC{p},\VEC{k}-\VEC{p})
        \nonumber \\
        && \hspace{1.2cm}
    + \left[ \VEC{k}\cdot\VEC{R}(\VEC{p}) \right]  B_{\rm N_{12}}(-\VEC{k},\VEC{k}-\VEC{p},\VEC{p}) \Big]\;.
\end{eqnarray}
In this equation, we can show
\begin{eqnarray}
   \int \frac{d^3p}{(2\pi)^3}
    \left(  \left[ \VEC{k}\cdot\VEC{R}(\VEC{k}-\VEC{p}) \right] 
    + \left[ \VEC{k}\cdot\VEC{R}(\VEC{p}) \right] \right)
     =  0\;,
\end{eqnarray}
and
\begin{eqnarray}
    && \int \frac{d^3p}{(2\pi)^3} \left[ \VEC{k}\cdot\VEC{R}(\VEC{p}) \right]P(\VEC{k}-\VEC{p}) \nonumber \\
    &=& 
    \int \frac{d^3p}{(2\pi)^3} \left[ \VEC{k}\cdot\VEC{R}(\VEC{k}-\VEC{p}) \right]P(\VEC{p})\;.
\end{eqnarray}
Therefore, we obtain
\begin{eqnarray}
    && P_{s}(\VEC{k})  \nonumber \\
    &=&
    \int \frac{d^3p}{(2\pi)^3}  \left[ \VEC{k}\cdot\VEC{R}(\VEC{p}) + \VEC{k}\cdot\VEC{R}(\VEC{k}-\VEC{p})\right]
    B(\VEC{k}-\VEC{p},-\VEC{k},\VEC{p}) 
    \nonumber \\
    &+& \frac{2}{\bar{n}}
    \int \frac{d^3p}{(2\pi)^3} \left[\VEC{k}\cdot\VEC{R}(\VEC{k}-\VEC{p})\right] P(\VEC{p}) \;.
    \label{Eq:P_rec_s1}
\end{eqnarray}

\subsubsection{$P_{\rm s^2}$}

Using Eq.~(\ref{Eq:T_NN}), the terms in $P_{s^2}$ consisting of the product of two power spectra can be calculated as
\begin{eqnarray}
    && P_{s^2, PP}(\VEC{k}) \nonumber \\
    &=& - P(\VEC{k}) \int \frac{d^3p}{(2\pi)^3}
    \left[ \VEC{k}\cdot\VEC{R}(\VEC{p}) \right]^2 P_{\rm N}(\VEC{p}) \nonumber \\
    &+&   \int \frac{d^3p}{(2\pi)^3}    
    \Bigg\{ \frac{1}{2} \left[ \VEC{k}\cdot\VEC{R}(\VEC{k}-\VEC{p}) \right]^2  P(\VEC{p})P_{\rm N}(\VEC{k}-\VEC{p})\nonumber \\
    &+&
      \frac{1}{2} \left[ \VEC{k}\cdot\VEC{R}(\VEC{p}) \right]^2  
    P(\VEC{k}-\VEC{p})P_{\rm N}(\VEC{p})  \nonumber \\
     &+&
    \left[ \VEC{k}\cdot\VEC{R}(\VEC{k}-\VEC{p}) \right] \left[ \VEC{k}\cdot\VEC{R}(\VEC{p}) \right]
    P_{\rm N}(\VEC{p}) P_{\rm N}(\VEC{k}-\VEC{p}) \Bigg\} \;. 
     \label{Eq:P_rec_s2_G}
\end{eqnarray}
Furthermore, the terms arising from the trispectrum are given by
\begin{eqnarray}
    && P_{s^2, T}(\VEC{k}) \nonumber \\
    &=&\int \frac{d^3p_1}{(2\pi)^3}\int \frac{d^3p_2}{(2\pi)^3}
    \left[ \VEC{k}\cdot\VEC{R}(\VEC{p}_1) \right]
    \left[ \VEC{k}\cdot\VEC{R}(\VEC{p}_2) \right] \nonumber \\
    &\times&
    \Big\{
        T_{\rm N_{12}}(\VEC{k}-\VEC{p}_1-\VEC{p}_2,-\VEC{k},\VEC{p}_1,\VEC{p}_2)\nonumber \\
    && - T_{\rm N_{12}}(\VEC{k}-\VEC{p}_1,-\VEC{k}-\VEC{p}_2, \VEC{p}_1,\VEC{p}_2)
    \Big\} \;.
\end{eqnarray}
Since the trispectrum is of the 2-loop order in the power spectrum calculations, we ignore this term in this paper. 

\section{Perturbation Theory Approach}
\label{Sec:PT}

The results presented in Section~\ref{Sec:PostRecon_P} hold in general when the post-reconstruction power spectrum is expanded to the second-order of $\VEC{s}$. In this subsection, we calculate the power spectrum and the bispectrum that appear in Section~\ref{Sec:PostRecon_P} using perturbation theory.

\subsection{Density fluctuations}

The $n$th-order of the pre-reconstruction galaxy density fluctuation is represented as
\begin{eqnarray}
    \widetilde{\delta}_n(\VEC{k})
    &=& \int \frac{d^3p_1}{(2\pi)^3} \cdots\int \frac{d^3p_n}{(2\pi)^3}
    (2\pi)^3\delta_{\rm D}(\VEC{k} - \VEC{p}_{[1,n]}) \nonumber \\
    &\times& 
    Z_n(\VEC{p}_1,\cdots,\VEC{p}_n)\, \widetilde{\delta}_{\rm lin}(\VEC{p}_1)\cdots 
    \widetilde{\delta}_{\rm lin}(\VEC{p}_n)\;,
\end{eqnarray}
where $\VEC{p}_{[1,n]}=\VEC{p}_1+\cdots+\VEC{p}_n$, $\delta_{\rm lin}$ means the linear dark matter density fluctuation. The kernel function $Z_n$ includes the RSD and bias effects, and it depends on the unit vector $\hat{n}$ in the line-of-sight direction since it includes the RSD effect, but we omit it here for notational simplicity. The first-order kernel function is given by
\begin{eqnarray}
    Z_1(\VEC{k}) = b_1 + f\, \mu^2\;,
\end{eqnarray}
where $\mu=\hat{k}\cdot\hat{n}$, $b_1$ is the linear bias parameter, and $f$ is the linear growth rate function. For the specific forms of $Z_{n\geq2}$, see, for example, Appendix C of~\citet{Sugiyama:2024eye}.

\subsection{1-loop corrections}

In the context of cosmological perturbation theory, the leading-order solution is referred to as the tree-level, and the next-leading order solution as the 1-loop level. The tree-level power spectrum is expressed using the first-order kernel function $Z_1$. Furthermore, in linear theory, the power spectrum does not change before and after reconstruction, hence we obtain 
\begin{eqnarray}
    P_{\rm rec,tree}(\VEC{k}) = P_{\rm tree}(\VEC{k}) = Z^2_1(\VEC{k}) P_{\rm lin}(k)\;,
    \label{Eq:P_tree}
\end{eqnarray}
where $P_{\rm lin}$ is the linear dark matter power spectrum, given by
\begin{eqnarray}
    \langle \widetilde{\delta}_{\rm lin}(\VEC{k})\widetilde{\delta}_{\rm lin}(\VEC{k}') \rangle
    = (2\pi)^3\delta_{\rm D}(\VEC{k}+\VEC{k}') P_{\rm lin}(k)\;.
\end{eqnarray}

The non-linear correction term for the pre-reconstruction power spectrum at the 1-loop level is represented using $Z_2$ and $Z_3$ as follows:
\begin{eqnarray}
    P_{\rm 1\mathchar`-loop}(\VEC{k}) = P_{22}(\VEC{k}) + P_{13}(\VEC{k}) \;,
\end{eqnarray}
where
\begin{eqnarray}
    \hspace{-0.5cm}&& P_{22}(\VEC{k}) \nonumber \\
    \hspace{-0.5cm}&=& 2 \int \frac{d^3p}{(2\pi)^3}
    \left[ Z_2(\VEC{k}-\VEC{p},\VEC{p}) \right]^2 P_{\rm lin}(|\VEC{k}-\VEC{p}|) P_{\rm lin}(p)\;, 
    \label{Eq:P22}
\end{eqnarray}
and
\begin{eqnarray}
    && P_{13}(\VEC{k}) \nonumber \\
    &=& 6\, Z_1(\VEC{k})P_{\rm lin}(k)\,\int \frac{d^3p}{(2\pi)^3}
    Z_3(\VEC{k},\VEC{p},-\VEC{p})  P_{\rm lin}(p) \;.
    \label{Eq:P13}
\end{eqnarray}
In addition, the pre-reconstruction bispectrum at the tree-level is given by
\begin{eqnarray}
    && B_{\rm tree}(\VEC{k}_1,\VEC{k}_2,\VEC{k}_3) \nonumber \\
    &=& 
    2\, Z_2(\VEC{k}_1,\VEC{k}_2)Z_1(\VEC{k}_1) Z_1(\VEC{k}_2) P_{\rm lin}(k_1) P_{\rm lin}(k_2) \nonumber \\
    &&+ \mbox{2 perms.}
    \label{Eq:B_tree}
\end{eqnarray}

By substituting the tree-level power spectrum and bispectrum given in Eqs.~(\ref{Eq:P_tree}) and (\ref{Eq:B_tree}) into Eqs.~(\ref{Eq:P_rec_s1}) and (\ref{Eq:P_rec_s2_G}), we obtain the post-reconstruction 1-loop correction for the power spectrum:
\begin{eqnarray}
    P_{\rm rec,1\mathchar`-loop}(\VEC{k})
    = P_{\rm rec,22}(\VEC{k}) + P_{\rm rec,13}(\VEC{k})\;,
\end{eqnarray}
where
\begin{widetext}
    \begin{eqnarray}
    P_{\rm rec, 22}(\VEC{k})
    &=& P_{22}(\VEC{k}) \nonumber \\
    &+& 
     2\, \int \frac{d^3p}{(2\pi)^3}
     \left[ \VEC{k}\cdot\VEC{R}(\VEC{p}) +   \VEC{k}\cdot\VEC{R}(\VEC{k}-\VEC{p})\right] 
     Z_2(\VEC{k}-\VEC{p},\VEC{p})Z_1(\VEC{k}-\VEC{p})Z_1(\VEC{p})
        P_{\rm lin}(|\VEC{k}-\VEC{p}|) P_{\rm lin}(p)  \nonumber \\
    &+& 
    \frac{2}{\bar{n}}  \int \frac{d^3p}{(2\pi)^3}
     \left[ \VEC{k}\cdot\VEC{R}(\VEC{k}-\VEC{p}) \right] Z_1^2(\VEC{p})P_{\rm lin}(p)  \nonumber \\
    &+&  \frac{1}{2} \int \frac{d^3p}{(2\pi)^3}
   \left[ \VEC{k}\cdot\VEC{R}(\VEC{k}-\VEC{p}) \right]^2 Z_1^2(\VEC{p})P_{\rm lin}(p)
    \left[ Z_1^2(\VEC{k}-\VEC{p}) P_{\rm lin}(|\VEC{k}-\VEC{p}|) + \frac{1}{\bar{n}} \right] \nonumber \\
    &+&  \frac{1}{2} \int \frac{d^3p}{(2\pi)^3}
    \left[ \VEC{k}\cdot\VEC{R}(\VEC{p}) \right]^2 Z_1^2(\VEC{k}-\VEC{p})P_{\rm lin}(|\VEC{k}-\VEC{p}|)
    \left[ Z_1^2(\VEC{p}) P_{\rm lin}(p) + \frac{1}{\bar{n}} \right] \nonumber \\
    &+&  \int \frac{d^3p}{(2\pi)^3}
    \left[ \VEC{k}\cdot\VEC{R}(\VEC{p}) \right]
    \left[ \VEC{k}\cdot\VEC{R}(\VEC{k}-\VEC{p}) \right]
    \left[ Z_1^2(\VEC{p}) P_{\rm lin}(p) + \frac{1}{\bar{n}} \right]
    \left[ Z_1^2(\VEC{k}-\VEC{p})P_{\rm lin}(|\VEC{k}-\VEC{p}|) + \frac{1}{\bar{n}}\right] \;, \nonumber \\
   P_{\rm rec,13}(\VEC{k}) &=& P_{13}(\VEC{k}) \nonumber \\
   &+&
    4\,Z_1(\VEC{k})P_{\rm lin}(k)  \int \frac{d^3p}{(2\pi)^3}
    \left[ \VEC{k}\cdot\VEC{R}(\VEC{p})  +  \VEC{k}\cdot\VEC{R}(\VEC{k}-\VEC{p}) \right] 
    Z_2(-\VEC{k},\VEC{p})  Z_1(\VEC{p})P_{\rm lin}(p) 
   \nonumber \\
   &-& Z_1^2(\VEC{k}) P_{\rm lin}(k)  \int \frac{d^3p}{(2\pi)^3}
   \left[ \VEC{k}\cdot\VEC{R}(\VEC{p}) \right]^2
   \left[ Z_1^2(\VEC{p}) P_{\rm lin}(p) + \frac{1}{\bar{n}} \right]\;.
  \label{Eq:main1}
\end{eqnarray}
\end{widetext}
This is the first main result of this paper. When we take the $\bar{n}\to\infty$ limit and ignore the bias effects, Eq.~(\ref{Eq:main1}) is consistent with the expression for the post-reconstruction 1-loop corrections of the dark matter power spectrum, given by~\citet{Hikage:2017tmm,Hikage:2019ihj}.

\begin{figure*}
    \centering
    \includegraphics[width=\textwidth]{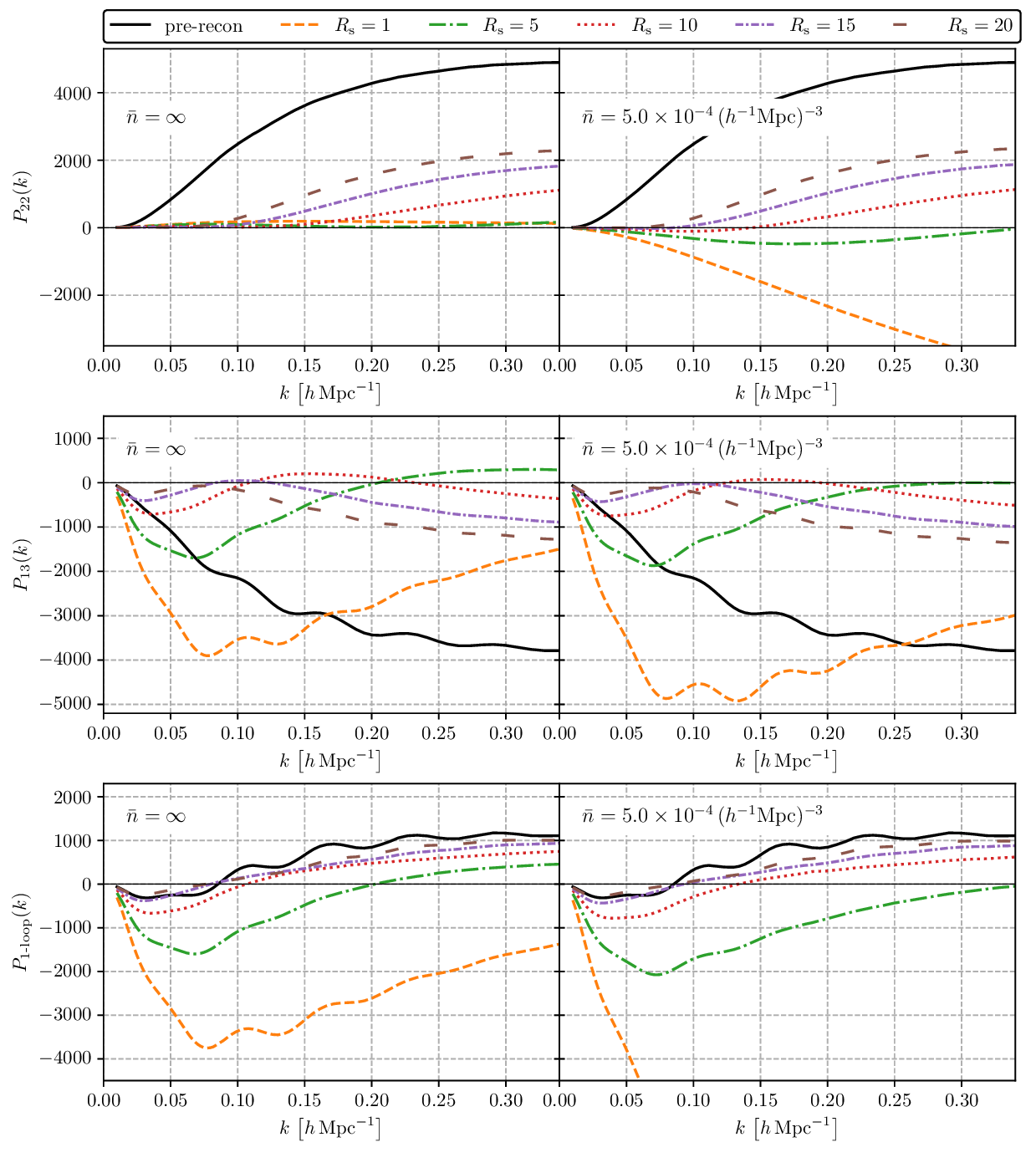}
    \caption{
        Upper panels: $P_{22}$ before and after reconstruction in real space at $z=0.5$. The linear bias parameter is set to $b_1=b_{1,\,{\rm fid}}=2$,and the nonlinear bias effect is neglected. The right panel shows the case with discreteness effects $\bar{n}=5.0\times10^{-4}\, h^{3}{\rm Mpc}^{-3}$, while the left panel shows the case without discreteness ($\bar{n}=\infty$). The black lines represent the pre-reconstruction results, and the post-reconstruction results are shown for different values of the input smoothing parameter: $R_{\rm s}=1$ (orange), $5$ (green), $10$ (red), $15$ (purple), and $20\, h^{-1}{\rm Mpc}$ (brown). Middle panels: Same as the top panels, but for $P_{13}$. Bottom panels: Same settings as the top panels, but showing the results for $P_{1\mathchar`-{\rm loop}}=P_{22}+P_{13}$.}
    \label{fig:P}
\end{figure*}


\begin{figure}[t]
    \centering
    \includegraphics[width=\columnwidth]{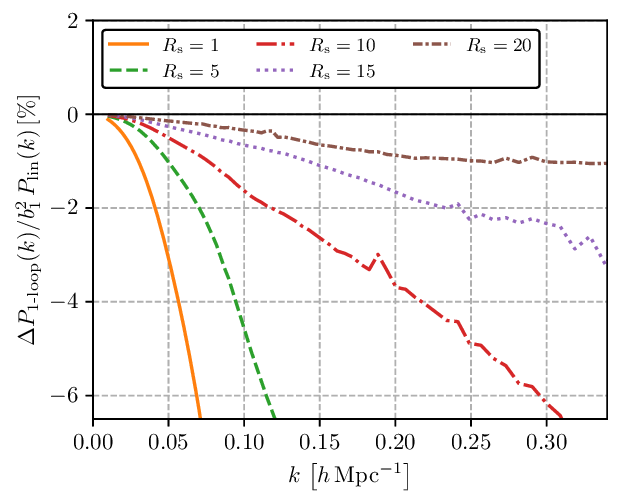}
    \caption{
        The ratio of the 1-loop correction difference with and without discreteness effects to the linear power spectrum in real space at $z=0.5$. The results are shown for five smoothing parameters: $R_{\rm s}=1$ (orange), $5$ (green), $10$ (red), $15$ (purple), and $20\,h^{-1}{\rm Mpc}$ (brown).
}
    \label{fig:P1loop}
\end{figure}

Figure~\ref{fig:P} compares the results for $P_{22}$, $P_{13}$, and $P_{1\mathchar`-{\rm loop}}$ in the presence and absence of reconstruction-specific discreteness effects. For simplicity, the solutions are plotted in real space (i.e., assuming $f=0$). The observational parameters are set to match a DESI-like survey, with $z=0.5$, $b_1=b_{1,\,{\rm fid}}=2$, and $\bar{n}=5.0\times10^{-4}\,h^3{\rm Mpc}^{-3}$. The left side of the top, middle, and bottom panels show results without discreteness effects (i.e., $\bar{n}=\infty$), which are comparable to the results of \citet{Hikage:2017tmm}, except for the bias effect.

Focusing on $P_{22}$, we find that without discreteness effects, the nonlinear contribution from $P_{22}$ decreases as the input smoothing parameter $R_{\rm s}$ becomes smaller. In contrast, when discreteness effects are included, decreasing $R_{\rm s}$ leads to poorer convergence of the wave number integral $\int d^3p$ in Eq.~(\ref{Eq:main1}), which in turn introduces spurious nonlinearities in $P_{22}$. This effect becomes particularly significant when $R_{\rm s}$ is less than $5\,h^{-1}{\rm Mpc}$. For example, see the dashed orange line for $R_{\rm s}=1\,h^{-1}{\rm Mpc}$ in the upper right panel. Similarly, for $P_{13}$, the discreteness effects become more pronounced for $R_{\rm s}=1$ and $5\,h^{-1}{\rm Mpc}$. Therefore, to effectively mitigate these effects, it is crucial to choose an appropriate value for $R_{\rm s}$, such as $R_{\rm s}=10$, $15$, or $20\,h^{-1}{\rm Mpc}$.

To highlight the importance of discreteness effects in accurate power spectrum modeling, Figure~\ref{fig:P1loop} shows the ratio of the 1-loop correction with and without discreteness (i.e., the difference between the right and left panels in the bottom row of Figure~\ref{fig:P}) normalized by the linear power spectrum. At $k=0.2\hk$, discreteness effects manifest at the level of about $4\%$ for $R_{\rm s}=10\hMpc$ and $2\%$ for $R_{\rm s}=15\hMpc$. Thus, we conclude that accounting for discreteness effects is essential for accurately modeling the post-reconstruction galaxy power spectrum at the percentage level.


\subsection{IR Limit}
\label{Sec:IRlimit}

The infrared (IR) effect refers to the non-linear effects that arise from scales significantly larger than the scale of interest $k$, namely, from smaller wave numbers $p$. At the 1-loop level, it is possible to extract only the IR effect by taking the limit of $p/k\to0$ in Eq.~(\ref{Eq:main1}), which is known as the solution in the IR limit or high-$k$ limit. In the pre-reconstruction power spectrum, the 1-loop correction terms, given by $P_{22}$ and $P_{13}$, cancel out in the IR limit~\cite{Suto:1990wf,Makino:1991rp}. This IR cancellation is known to hold at any order in perturbation theory~\citep{Jain:1995kx,Scoccimarro:1995if,Kehagias:2013yd,Peloso:2013zw,Sugiyama:2013pwa,Sugiyama:2013gza,Blas:2013bpa,Blas:2015qsi,Lewandowski:2017kes}. 

Furthermore, this non-perturbative IR cancellation has been shown to remain effective even after reconstruction~\cite{Sugiyama:2024eye}. However, this previous study on reconstruction did not take into account the discrete effects specific to reconstruction. In this section, as a preparation for the construction of the IR-resummed model in Section~\ref{Sec:IRmodel}, we compute the 1-loop correction term in the IR limit when discrete effects are included after reconstruction and show the occurrence of IR cancellation.

In Eqs.~(\ref{Eq:P22}), (\ref{Eq:P13}), and (\ref{Eq:main1}), we assume
\begin{eqnarray}
    Z_1(\VEC{k}-\VEC{p}) &\xrightarrow[p/k\to0]{}& Z_1(\VEC{k}) \;, \nonumber \\
    P_{\rm lin}(|\VEC{k}-\VEC{p}|) &\xrightarrow[p/k\to0]{}& P_{\rm lin}(k) \;.
\end{eqnarray}
In addition, the non-linear kernel functions are related to lower-order functions in the IR limit as follows:
\begin{eqnarray}
    Z_2(\VEC{k}-\VEC{p},\VEC{p}) 
    &\xrightarrow[p/k\to0]{}&
    \frac{1}{2} \left(  \frac{\VEC{k}\cdot\VEC{M}\cdot\VEC{p}}{p^2}  \right)Z_1(\VEC{k})\;, \nonumber \\
    Z_2(-\VEC{k},\VEC{p}) 
    &\xrightarrow[p/k\to0]{}&
    -\frac{1}{2} \left(  \frac{\VEC{k}\cdot\VEC{M}\cdot\VEC{p}}{p^2}  \right)Z_1(\VEC{k})\;, \nonumber \\
    Z_3(\VEC{k},\VEC{p},-\VEC{p}) 
    &\xrightarrow[p/k\to0]{}&
    -\frac{1}{3!} \left(  \frac{\VEC{k}\cdot\VEC{M}\cdot\VEC{p}}{p^2}  \right)^2Z_1(\VEC{k})\;,
\end{eqnarray}
where $\VEC{M}$ is the transformation matrix from real-space to redshift-space in linear Lagrangian perturbation theory, given by~\cite{Matsubara:2007wj}
\begin{eqnarray}
    M_{ij} = \delta_{ij} + f\hat{n}_i\hat{n}_j \;.
\end{eqnarray}
Note that since the dependence between $\VEC{p}$ and $\VEC{k}-\VEC{p}$ is symmetric, the variable transformation 
$\VEC{k}-\VEC{p}=\VEC{p}'$ and the limit of $p'/k\to 0$ should be computed in the same way.

Leaving only terms of the order of $(k/p)^2$ in the IR limit calculation, the post-reconstruction 1-loop correction terms are calculated as
\begin{eqnarray}
    P_{\rm rec,22,IR}(\VEC{k}) = X(\VEC{k})Z_1^2(\VEC{k}) P_{\rm lin}(k) \;, \nonumber \\
    P_{\rm rec,13,IR}(\VEC{k}) = - X(\VEC{k})Z_1^2(\VEC{k}) P_{\rm lin}(k) \;,
    \label{Eq:P_22_13_IR}
\end{eqnarray}
where the subscript ``IR'' means the value estimated in the IR limit. The function $X(\VEC{k})$ is given by
\begin{eqnarray}
    X(\VEC{k})
    &=& 
    \int \frac{d^3p}{(2\pi)^3}
    \left[ \left(  \frac{\VEC{k}\cdot\VEC{M}\cdot\VEC{p}}{p^2}  \right)
    +  
    \left[ \VEC{k}\cdot\VEC{R}(\VEC{p}) \right] Z_1(\VEC{p}) \right]^2 P_{\rm lin}(p) \nonumber \\
    &+& 
     \frac{1}{\bar{n}}
    \int \frac{d^3p}{(2\pi)^3}
    \left[ \VEC{k}\cdot\VEC{R}(\VEC{p}) \right]^2 \;.
\end{eqnarray}
Eq.~(\ref{Eq:P_22_13_IR}) shows that the post-reconstruction 1-loop correction terms cancel each other out in the IR limit even after accounting for the discrete effect, resulting in
\begin{eqnarray}
    P_{\rm rec,1\mathchar`-loop,IR}(\VEC{k}) = P_{\rm rec,22,IR}(\VEC{k}) + P_{\rm rec,13,IR}(\VEC{k}) 
    =  0\;.
\end{eqnarray}

The function $X(\VEC{k})$ can be further expressed as 
\begin{eqnarray}
    X(\VEC{k}) =  k^2 \sigma_{\perp}^2 ( 1-\mu^2 ) + k^2 \sigma_{\parallel}^2 \mu^2\;.
\end{eqnarray}
Here, $\sigma_{\perp}^2$ and $\sigma_{\parallel}^2$ are decomposed into three components,
\begin{eqnarray}
    \sigma_{\perp}^2 &=&  \sigma_{\rm pp\,,\perp}^2 + \sigma_{\rm ps\,,\perp}^2 + \sigma_{\rm ss\,,\perp}^2
    \nonumber \\
    \sigma_{\parallel}^2 &=& \sigma_{\rm pp\,,\parallel}^2 
    + \sigma_{\rm ps\,,\parallel}^2 + \sigma_{\rm ss\,,\parallel}^2\;,
    \label{Eq:Sigma}
\end{eqnarray}
where the subscript ``pp'' denotes the auto-correlation of the dark matter displacement vector $\PP$, ``ps'' denotes the cross-correlation between $\PP$ and the displacement vector $\VEC{s}$ for reconstruction, and ``ss'' denotes the auto-correlation of $\VEC{s}$. Each of these components has the following specific forms~\cite{Sugiyama:2024eye}:
\begin{widetext}
    \begin{eqnarray}
    \sigma^2_{\rm pp,\perp} &=&
    \frac{1}{3}\int \frac{dp}{2\pi^2} P_{\rm lin}(p), \nonumber \\
    \sigma^2_{\rm pp,\parallel} &=& (1+f)^2\, \sigma^2_{\perp}, \nonumber \\
        \sigma^2_{\rm ps, \perp}
        &=& \frac{1}{3} \int \frac{dp}{2\pi^2} \left( -\frac{W_{\rm G}(pR_{\rm s})}{b_{1,\rm fid}} \right)  
        \left[ 2\left( b_1+\frac{f}{5} \right)    \right]P_{\rm lin}(p)\;, \nonumber \\
        \sigma^2_{\rm ps, \parallel}
        &=& \frac{1}{3} \int \frac{dp}{2\pi^2} \left( -\frac{W_{\rm G}(pR_{\rm s})}{b_{1,\rm fid}} \right)
        \Bigg[ 2\left( 1+f \right)\left(  b_1+\frac{3}{5}f\right)
              \Bigg]P_{\rm lin}(p)\;, \nonumber \\
        \sigma^2_{\rm ss, \perp}
        &=& \frac{1}{3} \int \frac{dp}{2\pi^2} \left( -\frac{W_{\rm G}(pR_{\rm s})}{b_{1,\rm fid}} \right)^2
        \left[ \left( b_1^2+\frac{2}{5}b_1f + \frac{3}{35}f^2 \right) P_{\rm lin}(p) + \frac{1}{\bar{n}}   \right]\;, \nonumber \\
        \sigma^2_{\rm ss, \parallel}
        &=& \frac{1}{3} \int \frac{dp}{2\pi^2} \left( -\frac{W_{\rm G}(pR_{\rm s})}{b_{1,\rm fid}} \right)^2 
        \Bigg[ \left( b_1^2 + \frac{42}{35}b_1f +\frac{3}{7}f^2  \right) P_{\rm lin}(p) + \frac{1}{\bar{n}}
        \Bigg]\;.
        \label{Eq:ps_ss}
    \end{eqnarray}
\end{widetext}

\section{IR-Resummed Model}
\label{Sec:IRmodel}

In this section, we examine the discrete effects within the IR-resummed model for the post-reconstruction power spectrum, as proposed by~\citet{Sugiyama:2024eye} (see also~\citet{Chen:2024tfp}). This model incorporates the resummation of IR effects and accounts for the 1-loop correction terms in the post-reconstruction power spectrum.

\subsection{IR-resummed model construction}


In this subsection, we give an overview of how to construct the IR-resummed power spectrum model. For a detailed discussion, see~\cite{Sugiyama:2013gza,Senatore:2014via,Baldauf:2015xfa,Blas:2016sfa,Senatore:2017pbn,Ivanov:2018gjr,Lewandowski:2018ywf,Sugiyama:2020uil,Sugiyama:2024eye}. To simplify the explanation, we focus on the case of dark matter in real space, though the same approach applies to galaxies in redshift space.

The non-linear power spectrum can be generally decomposed into a term proportional to the linear dark matter power spectrum and a term consisting of mode coupling integrals~\citep{Crocce:2005xy,Crocce:2007dt}:
\begin{eqnarray}
    P(\VEC{k}) = G^2(\VEC{k})P_{\rm lin}(k) + P_{\rm MC}(\VEC{k}) \;,
    \label{Eq:P_G_MC}
\end{eqnarray}
where $G$ is called the propagator and $P_{\rm MC}$ is the mode-coupling term.

In the IR limit, the propagator and the mode-coupling terms are calculated as
\begin{eqnarray}
    G^2(\VEC{k})P_{\rm lin}(k) &=& e^{-k^2\sigma^2}P_{\rm lin}(k) \;, \nonumber \\
    P_{\rm MC}(\VEC{k}) &=& \left( 1 -  e^{-k^2\sigma^2}\right)\, P_{\rm lin}(k) \;,
    \label{Eq:P_G_MC_IR}
\end{eqnarray}
where $\sigma^2 = \sigma^2_{\rm pp,\,\perp}$ is given in Eq.~(\ref{Eq:ps_ss}). At the SPT 1-loop level, $P_{13}$ corresponds to the propagator term $G^2P_{\rm lin}$, while $P_{22}$ corresponds to the mode-coupling term $P_{\rm MC}$. Expanding $e^{-k^2\sigma^2}$, we find that $P_{13} = -k^2\sigma^2P_{\rm lin}$ and $P_{22}=k^2\sigma^2P_{\rm lin}$. Substituting Eq.~(\ref{Eq:P_G_MC_IR}) into Eq.~(\ref{Eq:P_G_MC}), the non-linearities cancel in the IR limit, leaving only the linear power spectrum:
\begin{eqnarray}
    P(\VEC{k}) &=& e^{-k^2\sigma^2}P_{\rm lin}(k) + \left( 1 -  e^{-k^2\sigma^2}\right)\, P_{\rm lin}(k)  \nonumber \\
    &=&P_{\rm lin}(k)\;.
\end{eqnarray}

The $P_{\rm MC}$ in the IR limit is proportional to $P_{\rm lin}$ and has the BAO signal, but the actual behavior of the BAO within $P_{\rm MC}$ is negligible. For example, the $P_{22}$ term is known to have smooth curves in numerical calculations (e.g., see Figure~\ref{fig:P}). Based on these numerical and experimental findings, we apply a procedure to remove the BAO signal from the mode coupling term in the IR limit. To achieve this, we decompose the linear matter power spectrum into a ``wiggle'' part ($P_{\rm w}$) containing only the BAO signal, and a ``no-wiggle'' part ($P_{\rm nw}$) without the BAO signal: $P_{\rm lin}(k) = P_{\rm w}(k) + P_{\rm nw}(k)$~\cite{Eisenstein:1997ik,Hamann:2010pw,Chudaykin:2020aoj}. We then replace $P_{\rm lin}$ in $P_{\rm MC}$ with $P_{\rm nw}$ and construct the no-wiggle mode-coupling term:
\begin{eqnarray}
    P_{\rm MC,\,nw}(k) = \left( 1 - e^{-k^2\sigma^2} \right)P_{\rm nw}(k)\;.
\end{eqnarray}
Using this no-wiggle mode-coupling term, we obtain the IR-resummed model at the linear level,
\begin{eqnarray}
    P(k) &=& e^{-k^2\sigma^2}P_{\rm lin}(k) + \left( 1 -   e^{-k^2\sigma^2}\right)P_{\rm nw}(k) \nonumber \\
    &=&P_{\rm nw}(k) +  e^{-k^2\sigma^2} P_{\rm w}(k)  
    \label{Eq:BAO_ap}
\end{eqnarray}
This method of constructing an IR-resummed power spectrum model was first proposed by~\citet{Sugiyama:2013gza}.

Subsequent studies~\cite{Baldauf:2015xfa,Vlah:2015zda,Senatore:2014via,Blas:2016sfa,Senatore:2017pbn,Ivanov:2018gjr,Lewandowski:2018ywf} have highlighted the significance of the wiggle part in mode couplings. By decomposing the mode-coupling term into its wiggle and no-wiggle components, we obtain the following expression in the IR limit:
\begin{eqnarray}
    P_{\rm MC}(k) &=& \left( 1 - e^{-k^2\sigma^2} \right)P_{\rm nw}(k) \nonumber \\
    &+& \left( e^{-k^2\sigma_{\rm BAO}^2} -  e^{-k^2\sigma^2}\right) P_{\rm w}(k)\;.
\end{eqnarray}
The parameter $\sigma_{\rm BAO}^2$ is given by
\begin{eqnarray}
    \sigma_{\rm BAO}^2 = \frac{1}{3}\int \frac{dp}{2\pi^2} \left( 1 - j_0(pr_{\rm BAO}) \right)P_{\rm lin}(p),
\end{eqnarray}
where $j_0$ is the $0$th order spherical Bessel function and $r_{\rm BAO}\sim110\hMpc$ is the BAO scale. A version including $j_2$ also exists, but it is not included here. This expression shows that the wiggle part from the propagator term cancels out completely with that from the mode-coupling term, and the remaining factor $e^{-k^2\sigma_{\rm BAO}^2}$ describes the BAO non-linearity. Thus, the final expression for the power spectrum becomes
\begin{eqnarray}
    P(k) &=& e^{-k^2\sigma^2}\left( P_{\rm w}(k) + P_{\rm nw}(k) \right)+ \left( 1 - e^{-k^2\sigma^2} \right)P_{\rm nw}(k) \nonumber \\
    &+& \left(e^{-k^2\sigma_{\rm BAO}^2} - e^{-k^2\sigma^2} \right) P_{\rm w}(k) \nonumber \\
    &=&P_{\rm nw}(k) + e^{-k^2\sigma_{\rm BAO}^2}   P_{\rm w}(k)  \;.
    \label{Eq:BAO}
\end{eqnarray}
Eq.~(\ref{Eq:BAO_ap}) and Eq.~(\ref{Eq:BAO}) are similar, but the parameter $\sigma^2$ describing the BAO non-linearity in Eq.~(\ref{Eq:BAO_ap}) is replaced by $\sigma^2_{\rm BAO}$ in Eq.~(\ref{Eq:BAO}). However, \citet{Blas:2016sfa} point out that in a $\Lambda$CDM model, the numerical difference between $\sigma_{\rm BAO}^2$ and $\sigma^2$ is small. In fact, \citet{Sugiyama:2024eye} shows that the difference between $\sigma_{\rm BAO}^2$ and $\sigma^2$ affects the power spectrum by about $0.2\%$, which is why the wiggle part in the mode-coupling term, $( e^{-k^2\sigma_{\rm BAO}^2} - e^{-k^2\sigma^2} )P_{\rm w}$, is negligible in numerical calculations, as seen in Figures~\ref{fig:P}. Hence, ignoring the wiggle part in the mode-coupling term, as in Eq.~(\ref{Eq:BAO_ap}), is a good approximation for constructing a model that fits the observed power spectrum.

In the post-reconstruction case, \citet{Sugiyama:2024eye} showed that the IR effect in the density fluctuations is mathematically similar to the pre-reconstruction case. This means that the method of constructing the pre-reconstruction IR-resummed model can be applied in the same way as in the post-reconstruction case. As a result, the non-linear damping of the BAO in the post-reconstruction power spectrum can be described by a single Gaussian damping function, as in the pre-reconstruction case~\cite{Sugiyama:2024eye,Chen:2024tfp}. It has long been thought that the post-reconstruction BAO behavior can be described by two exponential damping functions~\cite{Padmanabhan:2008dd,White:2010qd,Seo:2015eyw,White:2015eaa,Chen:2019lpf}, based on the Zel'dovich approximation (ZA)~\cite{Zeldovich:1969sb}. However, \citet{Sugiyama:2024eye} showed that ZA cannot account for all post-reconstruction IR effects. These theoretical developments have provided the theoretical basis for the recent BAO analysis performed by the DESI project~\cite{DESI:2024mwx}.

Furthermore, \citet{Sugiyama:2024eye} finds that the difference between $\sigma_{\rm BAO}^2$ and $\sigma^2$ after reconstruction is significantly smaller than before reconstruction. The post-reconstruction $\sigma_{\rm BAO}^2$ is given by
\begin{eqnarray}
    \hspace{-0.5cm}&& \sigma_{\rm BAO,\,rec}^2 \nonumber \\
    \hspace{-0.5cm} &=& \frac{1}{3}\int \frac{dp}{2\pi^2} \left( 1 - W(pR_{\rm s}) \right)^2 \left( 1 - j_0(pr_{\rm BAO}) \right) P_{\rm lin}(p)\;.
\end{eqnarray}
As $p\to0$, the factor $(1-W(pR_{\rm s}))^2$ approaches zero, suppressing the contribution from large scales. This reduces the large-scale contribution around $r_{\rm BAO}\sim110\hMpc$, meaning the impact of $j_0(pr_{\rm BAO})$ becomes negligible. In fact, \citet{Sugiyama:2024eye} shows that when $R_{\rm s}=15\hMpc$,  the impact of the difference between $\sigma_{\rm BAO}^2$ and $\sigma^2$ on the post-reconstruction power spectrum is extremely small, on the order of $0.0005\%$. This result demonstrates that the approximation of neglecting the wiggle part in the mode-coupling term holds very well after reconstruction.


\subsection{IR-resummed model at the 1-loop level}


Based on the discussion in the previous subsection, we construct an IR-resummed model of the post-reconstruction galaxy power spectrum at the 1-loop level. This model equivalent to that presented by~\citet{Sugiyama:2024eye}, except that it includes the discrete effects.

First, it was confirmed in Section~\ref{Sec:IRlimit} that all 1-loop correction terms in the IR limit cancel, even when discrete effects from the reconstruction are included. This implies that IR effects can be treated similarly, whether or not discrete effects are present. Therefore, the function ${\cal D}(\VEC{k}))$, representing the nonlinear decay of the BAO signal, is constructed as 
\begin{eqnarray}
    {\cal D}(\VEC{k}) &=&  \exp\left( -X(\VEC{k})/2 \right) \nonumber \\
    &=& \exp\left( - \frac{k^2 \sigma_{\perp}^2 ( 1-\mu^2 ) + k^2 \sigma_{\parallel}^2 \mu^2}{2} \right) \;,
    \label{Eq:D}
\end{eqnarray}
where the function $X(\VEC{k})$ is derived from the solution of the IR limit of $P_{22}$ or $P_{13}$ in Eq.~(\ref{Eq:P_22_13_IR}), and the specific expressions for $\sigma_{\perp}^2$ and $\sigma_{\parallel}^2$ are given in Eq.~(\ref{Eq:ps_ss}).

Next, at the 1-loop level, the propagator term and the mode-coupling term are given by~\cite{Sugiyama:2013gza,Sugiyama:2020uil,Sugiyama:2024eye}
\begin{eqnarray}
    && G^2(\VEC{k}) P_{\rm lin}(k) \nonumber \\
    &=& 
    {\cal D}^2(\VEC{k}) \left[ 1 - \ln {\cal D}^2(\VEC{k}) \right]Z_1^2(\VEC{k}) P_{\rm lin}(k) 
    \nonumber \\
    &+& {\cal D}^2(\VEC{k})P_{\rm rec,13}(\VEC{k}) \;,
    \label{Eq:G1loop}
\end{eqnarray}
and
\begin{eqnarray}
    && P_{\rm MC}(\VEC{k}) \nonumber\\
    &=& 
    \left( 1 - {\cal D}^2(\VEC{k}) \right)
    \left[ 1 - \ln {\cal D}^2(\VEC{k}) \right]Z_1^2(\VEC{k}) P_{\rm lin}(k) 
    \nonumber \\
    &+& \left( 1 - {\cal D}^2(\VEC{k}) \right) P_{\rm rec,13}(\VEC{k})\nonumber \\
    &+& \left[  P_{\rm rec,22}(\VEC{k}) + \ln {\cal D}^2(\VEC{k})Z_1^2(\VEC{k}) P_{\rm lin}(k)  \right]\;.
    \label{Eq:PMC1loop}
\end{eqnarray}
Substituting Eqs.~(\ref{Eq:G1loop}) and (\ref{Eq:PMC1loop}) into Eq.~(\ref{Eq:P_G_MC}), the IR effects cancel out, leading to the convergence of the 1-loop power spectrum in SPT. 

Third, we replace $P_{\rm lin}$ appearing in Eq.~(\ref{Eq:PMC1loop}) with $P_{\rm nw}$ and construct the no-wiggle mode-coupling term as
\begin{eqnarray}
    && P_{\rm MC,\, nw}(\VEC{k}) \nonumber\\
    &=& 
    \left( 1 - {\cal D}^2(\VEC{k}) \right)
    \left[ 1 - \ln {\cal D}^2(\VEC{k}) \right]Z_1^2(\VEC{k}) P_{\rm nw}(k) 
    \nonumber \\
    &+& \left( 1 - {\cal D}^2(\VEC{k}) \right) P_{\rm rec,13}(\VEC{k}) \left( P_{\rm nw}(k) / P_{\rm lin}(k) \right)\nonumber \\
    &+& \left[  P_{\rm rec,22}(\VEC{k}) + \ln {\cal D}^2(\VEC{k})Z_1^2(\VEC{k}) P_{\rm nw}(k)  \right]\;.
    \label{Eq:PMC1loop_nw}
\end{eqnarray}
Substituting Eqs.~(\ref{Eq:G1loop}) and (\ref{Eq:PMC1loop_nw}) into Eq.~(\ref{Eq:P_G_MC}), we finally obtain the IR-resummed model of the galaxy power spectrum at the 1-loop level:
\begin{eqnarray}
    P_{\rm rec}(\VEC{k}) 
    &=& {\cal D}^2(\VEC{k})
    \Big( \left[ 1 - \ln {\cal D}^2(\VEC{k}) \right]Z_1^2(\VEC{k}) \nonumber \\
    && \hspace{1cm}
    + \left[  P_{\rm rec,13}(\VEC{k})/ P_{\rm lin}(k)\right] \Big) P_{\rm w}(k)  \nonumber \\
    &+& Z_1^2(\VEC{k}) P_{\rm nw}(k) \nonumber \\
    &+& P_{\rm rec,22}(\VEC{k}) + \left[P_{\rm rec,13}(\VEC{k})/P_{\rm lin}(k)\right] P_{\rm nw}(k) \;.
    \label{Eq:main2}
\end{eqnarray}
This represents the second main result of this paper. In the ${\cal D}(\VEC{k})$ function, which represents the non-linear damping of BAO, the shot-noise terms included in the smoothing parameters $\sigma_{\rm ss,\perp}^2$ and $\sigma_{\rm ss,\parallel}^2$ are the same as those derived by~\citet{White:2010qd}. 

\begin{figure}[t]
    \centering
    \includegraphics[width=\columnwidth]{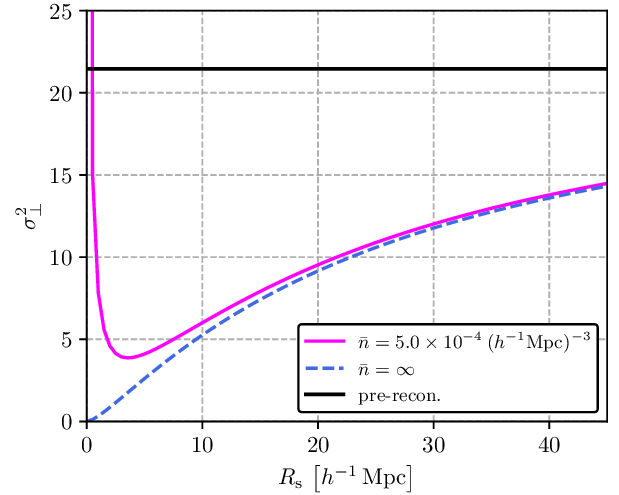}
    \caption{$\sigma_{\perp}^2$ in Eq.~(\ref{Eq:Sigma}) as a function of $R_{\rm s}$ in real space at $z=0.5$. The results are shown for the pre-reconstruction case (black), the post-reconstruction case without discrete effects (blue), and the post-reconstruction case with discrete effects (magenta).
}
    \label{fig:s2}
\end{figure}

\begin{figure}[t]
    \centering
    \includegraphics[width=\columnwidth]{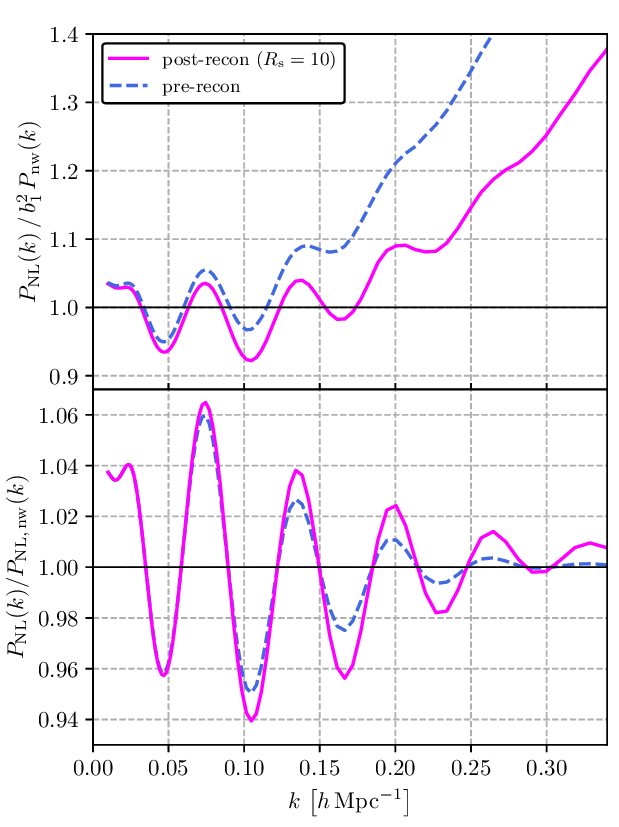}
    \caption{The IR-resummed power spectrum at $z=0.5$ in real space, including 1-loop corrections, is shown for the pre-reconstruction (blue) and post-reconstruction (magenta) cases. A smoothing parameter of $R_{\rm s}$ is used for the reconstruction. The top panel shows the results normalized to the linear no-wiggle power spectrum, while the bottom panel is normalized to the no-wiggle part of the nonlinear power spectrum.
}
    \label{fig:P_IR}
\end{figure}

Figure~\ref{fig:s2} shows the plot of $\sigma_{\perp}^2$ given by Eq.~(\ref{Eq:Sigma}) as a function of $R_{\rm s}$ in real space at $z=0.5$. The parameters $b_1=b_{1,\,{\rm fid}}=2$ and $\bar{n}=5\times10^{-4}\,h^3{\rm Mpc}^{-3}$, consistent with Figures~\ref{fig:P} and~\ref{fig:P1loop}, are used here. Since $\sigma_{\perp}^2=\sigma_{\parallel}^2$ in real space, it is sufficient to consider only $\sigma_{\perp}^2$. This parameter, which appears in Eq.~(\ref{Eq:D}), describes the Gaussian damping of the BAO signal. The smaller the value, the more the nonlinear effects on the BAO signal are suppressed, allowing for the recovery of linear BAO information.

In the absence of discrete reconstruction effects, $\sigma_{\perp}^2$ approaches zero as $R_{\rm s}$ decreases, indicating that the non-linear effects on the BAO signal disappear, as expected. However, for a realistic DESI-like survey with $\bar{n}=5\times10^{-4}\,h^3{\rm Mpc}^{-3}$, the wave number integral $\int d^3p$ in Eq.~(\ref{Eq:ps_ss}) becomes poorly convergent for $R_{\rm s}$ below about $5\hMpc$. Consequently, the value of $\sigma_{\perp}^2$ becomes extremely large. From this figure, it is evident that choosing $R_{\rm s}\geq10\hMpc$ is necessary to mitigate these discrete effects.

Figure~\ref{fig:P_IR} shows the IR-resummed power spectrum model, both before and after reconstruction, including the 1-loop corrections. The top panel shows the result normalized by the linear no-wiggle power spectrum, while the bottom panel shows the result normalized by the non-linear no-wiggle power spectrum, $P_{\rm NL,\,nw}= Z_1^2P_{\rm nw}+P_{ {\rm rec},22}+(P_{\rm nw}/P_{\rm lin})P_{ {\rm rec},13}$. As expected, the post-reconstruction IR-resummed model provided by Eq.~(\ref{Eq:main2}) successfully reduces the non-linear damping of the BAO signal and, even considering discrete effects, offers a correction to the shape of the power spectrum at the 1-loop level.


\section{Discussion and Conclusions}
\label{Sec:Conclusions}

This paper constructs a theoretical model of the post-reconstruction galaxy power spectrum in anticipation of future data analyses of the RSD effect using the galaxy distribution after reconstruction. To explain the observed galaxy power spectrum, it is necessary to consider various effects such as the non-linear gravitational effect, the RSD effect, and the bias effect. In particular, we point out the need to add shot noise terms due to discrete effects in the case of post-reconstruction analysis. Specifically, we present the shot noise terms that appear in the 1-loop corrections of the power spectrum in Standard Perturbation Theory (see Eq.~\ref{Eq:main1}). Furthermore, by applying this calculation to the resummation model of infrared (IR) effects, we show a power spectrum model applicable to small scales due to the 1-loop corrections, while describing the non-linear damping effects before and after the reconstruction of the BAO signal (see Eq.~\ref{Eq:main2}).

We have provided a discrete representation of the density fluctuations in the post-reconstruction galaxy distribution and calculated the associated power spectrum. Furthermore, we have demonstrated that the differences in the power spectrum before and after reconstruction can be described using higher-order statistics such as the bispectrum and trispectrum, in addition to the combination of the pre-reconstruction power spectra. These correction terms include shot-noise terms arising from discrete effects specific to the reconstruction. At the 1-loop level, contributions from the trispectrum can be ignored, yielding two types of terms in the post-reconstruction 1-loop power spectrum: one derived from the product of the pre-reconstruction power spectrum and another based on the bispectrum. 

The shot noise term appears alongside the Gaussian filtering function, $W_{\rm G}(pR_{\rm s}) = \exp\left( -p^2R_{\rm s}^2/2 \right)$, introduced for reconstruction. This suggests that when the smoothing scale $R_{\rm s}$ is small, the shape of the post-reconstruction power spectrum is significantly affected by the shot-noise term specific to the reconstruction. This effect is particularly pronounced when $R_{\rm s}$ is less than $5\hMpc$ (see Figures~\ref{fig:P} and \ref{fig:s2}). Even when $R_{\rm s}=10\hMpc$ or $15\hMpc$ is selected, Figure~\ref{fig:P1loop} indicates that there is a $2-4\%$ impact on the power spectrum at $k=0.2\hk$. In other words, it is crucial to properly account for the discrete effects in order to accurately model the post-reconstruction galaxy power spectrum.

Furthermore, we have derived the solution in the IR (high-$k$) limit of the post-reconstruction 1-loop power spectrum. The calculations in the IR limit are used to construct the IR-resummed model, and our calculations can reproduce the result shown by~\citet{White:2010qd}, i.e., the shot noise term that appears in the smoothing parameters that characterize the exponential damping function of the BAO signal.

The post-reconstruction power spectrum model presented in this paper includes almost all the effects that should be applied to actual observational data, such as gravitational non-linear effects, RSD effects, bias effects, exponential damping effects of BAO, and discrete effects unique to the reconstruction. Thus, it is immediately applicable to actual cosmological data analysis. Nonetheless, for more precise analyses, several improvements could be considered, summarized as follows:

\begin{enumerate}

    \item While our model incorporates both resummed IR effects and SPT 1-loop non-linear effects, considering higher order perturbation theory terms could extend its applicability to smaller scales.

    \item Given our findings, building emulators for the post-reconstruction power spectrum (e.g., see~\citet{Wang:2023hlx}) should be designed to account for changes in galaxy number densities due to the reconstruction-specific discrete effect. In other words, the galaxy number density input to the emulator should be set to exactly match the observed galaxy number density.

    \item The calculation of the displacement vector for reconstruction should include the window function effects of the observed region, necessitating a post-reconstruction power spectrum model that includes such window effects.

\end{enumerate}

\begin{acknowledgments}
N.S. acknowledges financial support from JSPS KAKENHI Grant No. 19K14703. 
N.S. thanks the referee for bringing up valuable comments and heavily improving the paper's quality.
\end{acknowledgments}

%
\bibliography{ms}

\end{document}